\documentclass[reprint,preprintnumbers,nofootinbib,amsmath,amssymb,floatfix,aps,longbibliography,prd]{revtex4-2}
\usepackage[utf8]{inputenc}
\usepackage{textcomp}
\usepackage{gensymb}
\usepackage{booktabs}
\usepackage{graphicx}
\usepackage{multirow}
\usepackage{hyperref}

\newcommand{\nuwro}{\textsc{NuWro}}
\newcommand{\MB}{MicroBooNE}
\newcommand{\MBC}{MicroBooNE Collaboration}
\newcommand{\ppi}{CC$1p0\pi$}
\newcommand{\etal}{{\it et al.}}

\newcommand{\ve}[1]{\ensuremath{\mathbf{#1}}}
\newcommand{\n}[1]{\ensuremath{|\mathbf{#1}|}}

\begin{document}
\title{JLab spectral functions of argon in NuWro and their implications for MicroBooNE}

\begin{abstract}
The Short-Baseline Neutrino program in Fermilab aims to resolve the nature of the low-energy excess events observed in LSND and MiniBooNE, and analyze with unprecedented precision neutrino interactions with argon. These studies require a~reliable estimate of neutrino cross sections, in particular for charged current quasielastic scattering (CCQE). Here, we report updates of the \nuwro{} Monte Carlo generator that, most notably, bring the state-of-the-art spectral functions to model the ground state properties of the argon nucleus, and improve the accuracy of the cross sections at low energies by accounting for the effects of the nuclear Coulomb potential. We discuss these developments in the context of electron and neutrino interactions, by comparing updated \nuwro{} predictions to experimental data from Jefferson Laboratory Hall A and MicroBooNE. The MicroBooNE CCQE-dominated data are described with the $\chi^2$ per degree of freedom of 0.7, compared with 1.0 in the local Fermi gas model. The largest improvement is observed for the angular distributions of the produced protons, where the $\chi^2$ reduces nearly by half. Being obtained using the axial form factor parametrization from MINERvA, our results indicate a~consistency between the CCQE measurements in MINERvA and MicroBooNE.
\end{abstract}

\author{Rwik Dharmapal Banerjee}
\author{Artur M. Ankowski}
\email{artur.ankowski@uwr.edu.pl}

\author{\\Krzysztof M. Graczyk}
\author{Beata E. Kowal}
\author{Hemant Prasad}
\author{Jan T. Sobczyk}

\affiliation{Institute of Theoretical Physics, University of Wroc\l aw, plac Maxa Borna 9,
50-204, Wroc\l aw, Poland}

\date{\today}

\maketitle

\section{Introduction}
Over the past 25 years, the phenomenon of neutrino oscillations has been extensively studied for a~broad range of energies and baselines, using accelerator, atmospheric, reactor, solar, and electron-capture neutrinos~\cite{Giganti:2017fhf}.
While most observations can be explained within the three-neutrino paradigm, this is not the case for the results of the gallium experiments~\cite{Elliott:2023cvh} and for the measurements performed by LSND~\cite{LSND:1997vun,LSND:2001aii} and MiniBooNE~\cite{MiniBooNE:2020pnu}.

LSND and MiniBooNE observed a~consistent excess~\cite{MiniBooNE:2020pnu} of electron (anti)neutrinolike charged-current quasielastic (CCQE) events appearing in muon (anti)neutrino beams. Due to the experimental limitations, those events could be composed of any number of $e^\pm$'s or  $\gamma$'s, and their origin remains a complete mystery~\cite{Abdullahi:2023ejc}.

To bring a definitive conclusion to the nature of the excess, the Short-Baseline Neutrino (SBN) program~\cite{MicroBooNE:2015bmn} in Fermilab is designed to perform precision measurements of neutrino oscillations at different baselines, using the Booster Neutrino Beam. The data from \MB{}~\cite{MicroBooNE:2016pwy}, already collected, will soon be complemented by those from the Short-Baseline Near Detector (SBND)~\cite{SBND:2020scp} and ICARUS~\cite{ICARUS:2023gpo}.

Collecting unprecedented event statistics will also enable cutting-edge precision studies of neutrino interactions, in preparation for the Deep Underground Neutrino Experiment~\cite{DUNE:2020lwj}.

The measurements performed in the SBN program require reliable estimates of neutrino cross sections. Particularly important are these for CCQE scattering, the dominant interaction mechanism at the SBN kinematics.

Here we report an improved description of neutrino interactions with argon in the \nuwro{} Monte Carlo generator. We implement the proton and neutron spectral functions (SFs) recently determined in a~coincidence electron-scattering experiment performed in Jefferson Laboratory (JLab) Hall A~\cite{JeffersonLabHallA:2022cit,JeffersonLabHallA:2022ljj}. We also update the \nuwro{} code to account for the distortion of the charged-lepton's kinematics induced by the Coulomb field of the nucleus~\cite{Aste:2004yz,Ankowski:2014yfa}, and for nuclear recoil. These effects play an important role at neutrino energies of a~few hundred MeV, where the excess appearance events are searched for in the SBN program. In addition, we add the axial form factor parametrizations from Refs.~\cite{Meyer:2016oeg,MINERvA:2023avz} and their uncertainties to the \nuwro{} code.

To assess the validity of the developed approach, we compare our predictions for inclusive electron scattering on argon to the experimental data from Ref.~\cite{Dai:2018gch}. As an illustrative example for neutrino interactions, we analyze various differential cross sections for an event class dubbed \ppi{} by the \MBC{}~\cite{MicroBooNE:2020fxd}, in which the CCQE contribution is enhanced.

For those experimental data, Franco-Patino \etal{}~\cite{Franco-Patino:2021yhd} showed that the predictions of the (global) Fermi gas model and the independent-particle shell model are in good agreement, but significantly overestimate the data. The root cause of this issue was not identified.

Butkevich~\cite{Butkevich:2021sfn} calculated the \ppi{} differential cross sections within the relativistic distorted-wave impulse
approximation (RDWIA) and found that they fit the \MB{} data with the $\chi^2$ per degree of freedom (d.o.f) of 1.12 (1.36) when the axial mass is set to 1.0 GeV (1.2 GeV)  in the dipole parametrization of the axial for factor.

In Ref.~\cite{Franco-Patino:2023msk}, Franco-Patino \etal{} compared the predictions of two RDWIA models and the SuSAv2 approach implemented in the {\sc genie} Monte Carlo generator to the \MB{} \ppi{} data. The authors of Ref.~\cite{Franco-Patino:2023msk} found the agreement to be the best for the RWDIA calculations based on a~phenomenological potential determined in proton-nucleus scattering, without any modifications, and the worst for the SuSAv2 approach in {\sc genie}.

This article is organized as follows. In Sec.~\ref{sec:NuWro} we summarize the features of \nuwro{} and present in detail the new developments. Our results are presented and discussed in Sec.~\ref{sec:results}. Finally, Sec.~\ref{sec:summary} summarizes our findings and describes the outlooks for future updates of \nuwro{}.

\section{NuWro Monte Carlo Generator}\label{sec:NuWro}
\nuwro{}~\cite{NuWro} is a~Monte Carlo generator of lepton-nucleus events in the few-GeV energy region, which has been developed primarily by the neutrino theory group in the University of Wroc{\l}aw since 2004~\cite{Sobczyk:2004va,Juszczak:2005wk}. Its principal application is in simulations for accelerator-based  neutrino experiments.

Interacting with nucleons, neutrinos in \nuwro{} can induce
\begin{itemize}
    \item[(i)] CCQE scattering, implemented according to the formalism of Llewellyn-Smith~\cite{LlewellynSmith:1971uhs},
    \item[(ii)] hyperon production~\cite{Thorpe:2020tym},
    \item[(iii)] single pion production, through the $\Delta$ resonance excitation or nonresonant pion production~\cite{Sobczyk:2004va},
    \item[(iv)] deep-inelastic scattering, described in the approach of Bodek and Yang~\cite{Bodek:2002vp}. It relies on leading-order parton-distribution functions~\cite{Gluck:1994uf} and extends the applicability of the parton model to the low-$Q^2$ region, $Q^2$ being the four-momentum squared, by incorporating higher-order corrections in effective target and final-state  masses.
\end{itemize}

From the set of options for the $\Delta$ form factors in \nuwro{}~\cite{Graczyk:2009qm,Lalakulich:2005cs,Barquilla-Cano:2007vds,Alvarez-Ruso:1997njo}, we choose those of Ref.~\cite{Graczyk:2009qm}. Given the small background of pion production, this choice has a negligible impact on the presented results.

Among the options available in \nuwro{} for describing nuclear effects in quasielastic scattering on the argon nucleus, we focus here on the SF approach and the local Fermi gas (LFG) model, employed by the \MBC{} in Ref.~\cite{MicroBooNE:2020fxd}.

In the SF approach, the shell structure and correlations between nucleons can be consistently taken into account in the local-density approximation~\cite{Benhar:1989aw,Benhar:1994hw}, by combining the shell-structure determined in coincidence electron-scattering experiments with the results of theoretical calculations for infinite nuclear matter at different densities. As such an experiment has only recently been performed for argon, until now \nuwro{} relied on an oversimplified estimate of the SFs to describe this nucleus~\cite{Ankowski:2007uy}. In this article, we present the implementation of the state-of-the-art knowledge of the ground state properties of argon~\cite{JeffersonLabHallA:2022cit,JeffersonLabHallA:2022ljj}, as detailed below.

In the LFG model, the shell structure of the nucleus and correlations between nucleons are both neglected. The nucleons bound in the nucleus are assumed to fully occupy the states with momenta up to the Fermi momentum, $p_F(r)$, which depends on the radial distance from the nuclear center, $r$, according to the nuclear density profile, $\rho(r)$. In \nuwro{}, $\rho(r)$ is approximated by the charge distribution~\cite{DeVries:1987atn}. Energy conservation is imposed as
\[
E_\nu + E_n -\epsilon(r)= E_\ell + E_p,
\]
where $E_\nu$, $E_\ell$, and $E_p$ are the energies of the neutrino, the charged lepton, and the produced proton. From the struck neutron's energy $E_n$, the separation energy depending on $r$ as
\[
\epsilon(r)=V+\sqrt{M^2+p_F^2(r)}
\]
is subtracted, with $V=7$ MeV and $M$ being the nucleon mass.

While in \nuwro{}, the SF approach can be employed to model nuclear effects in quasielastic scattering, the LFG model remains the underlying framework in other reaction channels. This treatment, admittedly, falls short of theoretical consistency. However, this issue has a minor impact on the numerical results presented here, in which CCQE interaction overwhelmingly dominates and other processes contribute only as minor backgrounds.

In addition to the reaction mechanisms described for scattering on nucleons, for atomic nuclei \nuwro{} also takes into account interactions involving two-body currents~\cite{Bonus:2020yrd}, commonly referred to as meson-exchange currents (MEC), and coherent pion production~\cite{Berger:2008xs}. Hadrons propagating in nuclear medium can produce intranuclear cascade, the modeling of which is discussed in Refs.~\cite{Golan:2012wx,Niewczas:2019fro}. 

As the interaction between a neutrino and a nucleon takes place inside the nucleus, energy exchanges with the spectator system lead to a more complicated energy conservation in the vertex than for a~free nucleon, resulting in a broadening of the double differential cross section. Additionally, when the struck nucleon propagates through the nucleus, its energy spectrum is changed by interactions with surrounding nucleons. These effects of final-state interactions (FSIs) in the argon nucleus are currently not accounted for. We leave for the future analyses their implementation in the framework of Ref.~\cite{Ankowski:2014yfa}.

\subsection{JLab spectral functions}
The experiment E12-14-012, performed in Hall A of JLab, in Newport News, Virginia, analyzed inclusive and exclusive electron scattering, using a 2.222-GeV beam. The inclusive data, in which only the scattered electron was detected, were collected for a variety of targets~\cite{Murphy:2019wed}, including argon~\cite{Dai:2018gch}.

The exclusive measurements, in which the scattered electron was detected in coincidence with a knocked-out proton, were conducted for argon~\cite{JeffersonLabHallA:2022cit} and titanium~\cite{JeffersonLabHallA:2022ljj}. The collected events were required to contain the proton in a~narrow opening angle along the direction of the momentum transferred by the electron to the nucleus, to reduce the effect of FSI.

The excitation energy of the residual nucleus was determined from the known energy transfer and the detected proton's energy. Similarly, the recoil momentum of the residual nucleus was obtained from the momentum transfer and the proton's momentum.

After correcting for FSI, the proton SFs $P^{T}_{(p)}(\ve p, E)$---the probabilities to remove a proton with momentum \ve p from the target nucleus $T$, leaving the residual system with excitation energy $E$---were determined for argon and titanium.

As the proton shell structure in titanium mirrors the neutron one in argon, the proton SF of titanium can be used as an approximation for the neutron SF of argon, with corrections to this picture coming, among others, from Coulomb interactions between protons. Note that currently, the neutron SFs cannot be determined experimentally in a direct manner.

In this analysis, we assume that
\begin{equation}
P^\text{Ar}_{(n)}(\ve p, E)=P^\text{Ti}_{(p)}(\ve p, E+S),
\end{equation}
with the value of the shift between the two SFs estimated to be $S=-1.6$ MeV,
based on the values of nuclear masses~\cite{Wang:2016}. We normalize the proton (neutron) SF of argon to 18 (22).

We account for the effect of Pauli blocking by the action of the step function, as in Ref.~\cite{Benhar:2005dj}, with the average Fermi momentum $\overline p_F=220$ MeV. We determine this value by weighting the local Fermi momenta with the density distribution unfolded from the measured charge density~\cite{DeVries:1987atn} according to Ref.~\cite{Kelly:2002if}.

Implementing the SFs in \nuwro{}, we do not treat separately their mean-field (MF) and correlated parts. The MF contribution describes the shell structure of the nucleus, and the correlated one gives the distribution of nucleons taking part in short-range interactions. When an event is generated using the correlated part of the SF, two nucleons should be involved in the primary interaction. The additional nucleon is currently not simulated by \nuwro{}.

Due to the cuts imposed in Ref.~\cite{MicroBooNE:2020fxd} and summarized in Appendix~\ref{sec:cuts}, the neutrino results presented in this article are only sensitive to additional protons with momentum above 300 MeV in the final state. Based on the SF alone, we estimate that such events cannot yield more than 12\% of the cross section, and nuclear transparency of argon can be expected to reduce this contribution by half. Further reduction is expected due to the higher energy transfers necessary to knock out the high-momentum correlated nucleons compared with those from the MF part of the SF. Therefore, we leave the improved implementation of the correlated SF for future updates.

To allow \nuwro{} users to gauge the effect that correlated nucleons may have on their simulations, we implement both the full and MF SFs. We use this option for comparisons to the \MB{} data that we discuss in Sec.~\ref{sec:results}.

\subsection{Low-energy improvements: The Coulomb effects}
The electrostatic potential of the nucleus alters the momenta of charged particles in its vicinity, increasing (decreasing) them for the negative (positive) charges. It also induces a focusing (defocusing) of their wave functions. These effects can be accurately accounted for in the effective momentum approximation, EMA$'$, introduced for electron scattering in Ref.~\cite{Aste:2004yz}.

In Ref.~\cite{Ankowski:2014yfa}, EMA$'$ was extended to CCQE interactions by modifying the charged-lepton's energy and momentum in the interaction vertex,
\begin{equation}\begin{split}
E_\ell&\rightarrow E_\ell^\textrm{eff}=E_\ell+sV_C,\\
\n{k_\ell}&\rightarrow\n{k_\ell^\textrm{eff}}=\sqrt{(E_\ell+sV_C)^2-m_\ell^2},
\end{split}\end{equation}
with $s=+1$ for neutrinos and $-1$ for antineutrinos, and by multiplying the cross section by the focusing factor
\begin{equation}
\left(\frac{\n{ k^\textrm{max}_{\ell}}}{\n{ k^\textrm{eff}_\ell}}\right)^2,
\end{equation}
where
\begin{equation}
\n{k_\ell^\textrm{max}}=\sqrt{(E_\ell+sV^\textrm{max}_C)^2-m_\ell^2}.
\end{equation} In the above equations, $m_\ell$ is the charged-lepton's mass $(\ell=e, \mu, \tau)$, $V_C$ the average Coulomb energy in the nucleus, and $V^\textrm{max}_C$ is the potential at the nuclear center.

We determine the Coulomb potential for argon using its measured charge density~\cite{DeVries:1987atn}. To calculate its average value, we weight it with the density distribution unfolded following to the procedure described in Ref.~\cite{Kelly:2002if}. The resulting values are $V^\textrm{max}_C= 9.6$ MeV and $V_C=7.3$ MeV.

The Coulomb effects play a significant role at low neutrino energies, but their influence diminishes as energies increase. The charged-lepton's energy spectrum undergoes a constant shift by $V_C$, which is sizable when a 200-MeV neutrino scatters on argon, but becomes much less noticeable when its energy is a few GeV.

This is also the case with the impact of the focusing factor. For example, for muon neutrinos of energy 200 MeV, the differential CCQE cross section $d\sigma/dE_\mu$ typically increases by 5\%--15\%, while at 600 MeV, the increase amounts to 1\%--3\% only.

The energy dependence of the Coulomb effects makes them of great importance in the search for low-energy excess events.

\subsection{Low-energy improvements: The nuclear recoil}
The updated SF implementation in \nuwro{} assumes energy conservation in the form
\[
E_\nu + M_A = E_\ell + E_p + \sqrt{(M_A-M+E)^2+\ve p^2},
\]
where $M_A$ is the target-nucleus mass, and $E$ is the excitation energy of the residual nucleus. The previously employed expression would correspond to vanishing recoil momentum $\ve p$.

Accounting for the recoil energy---remaining below 9~MeV and typically $\mathcal{O}$(1 MeV) for argon---results in a~somewhat reduced phase space that is accessible for the produced particles, the muon and the proton. This leads to a slight decrease in the CCQE cross sections.

For example, at energy 200 MeV, the differential CCQE cross section $d\sigma/dE_\mu$ for muon neutrinos typically decreases by 0.8\%--0.9\% and is shifted by 0.3--0.4 MeV, while at 600 MeV, this reduction becomes only 0.2\%, and the shift amounts to 0.2--0.3 MeV.

While the nuclear recoil does not modify the cross sections significantly, it is currently accounted for in \nuwro{} to avoid unnecessary approximations.

\subsection{Axial form factor}
Next to modeling of nuclear effects, the axial form factor $F_A(Q^2)$ is currently the most sizable source of uncertainties in the estimates of the CCQE cross sections for nuclear targets. Extractions of $F_A(Q^2)$ in neutrino experiments rely on flux determinations, background subtractions, and efficiency corrections, and are preferably performed for light targets---hydrogen and deuterium---where nuclear effects are not a~concern.

Until recently, the axial form factor was generally parametrized in the dipole form, with the $Q^2$ dependence driven by the axial mass $M_A$. In the literature, the $M_A$ value was frequently set to the world average of $1.026\pm0.021$ GeV, obtained by Bernard \etal~\cite{Bernard:2001rs}. The analysis presented in Ref.~\cite{Bernard:2001rs} performed averaging of the results of neutrino experiments performed between 1969 and 1990, including the axial mass extractions from charged- and neutral-current scattering of neutrinos and antineutrinos for a variety of targets, ranging from hydrogen and deuterium to freon and propane.

Intriguingly, various modern experiments~\cite{K2K:2006odf,MiniBooNE:2007iti,MiniBooNE:2010bsu,MiniBooNE:2010xqw,MINOS:2014axb} reported higher values of the axial mass, 1.20--1.35 GeV, albeit with large uncertainties, based on CCQE data for nuclear targets---oxygen, carbon, or iron.

In the context of \MB{}, the results from MiniBooNE are of particular importance, as they were collected using the same neutrino beam. In Ref.~\cite{MiniBooNE:2007iti}, the value $M_A=1.23\pm0.20$ GeV was determined, based on the event statistics higher by at least 2 orders of magnitude than that in the early experiments. A reanalysis of the MiniBooNE data, following an improved estimate of the CC single $\pi^+$ production background, brought an even higher value of the axial mass, $1.35\pm0.17$ GeV~\cite{MiniBooNE:2010bsu}, supported by a~subsequent neutral-current quasielastic study that found $M_A=1.35\pm0.11$ GeV~\cite{MiniBooNE:2010xqw}.

The past measurements of the differential cross sections $d\sigma/dQ^2$ for CCQE $\nu_\mu$ scattering on deuterium from Refs.~\cite{Baker:1981su,Miller:1982qi,Kitagaki:1983px} were reanalyzed by Meyer \etal{}~\cite{Meyer:2016oeg}, using updated vector form factors and accounting for uncertainties of experimental acceptance and nuclear effects. Contrary to the expectation based on the results of modern experiments~\cite{K2K:2006odf,MiniBooNE:2007iti,MiniBooNE:2010bsu,MiniBooNE:2010xqw,MINOS:2014axb}, the obtained axial form factor---the $z$-expansion fit with four free parameters---exhibits a~(somewhat) harder $Q^2$ dependence than that of Ref.~\cite{Bernard:2001rs}. The authors of Ref.~\cite{Meyer:2016oeg} also found the uncertainties of $F_A(Q^2)$ to be larger than those estimated by Bernard \etal~\cite{Bernard:2001rs}.

Due to safety concerns, currently there are no pro{\-}spects for measurements of CCQE scattering on a~hydrogen or deuterium targets, which could clarify the apparent tension between the old and new extractions of the axial form factor.

However, an innovative technique has been recently employed by the MINERvA experiment~\cite{MINERvA:2023avz}. From interactions of muon antineutrinos with a plastic scintillator target, CH, events on hydrogen have been selected by requiring the muon and neutron momenta perpendicular to the beam direction to balance each other. The measured CCQE cross section allowed the axial form factor to be determined free of nuclear effects. Within the uncertainties, the MINERvA result is in agreement with that of Bernard \etal~\cite{Bernard:2001rs} for $Q^2\lesssim0.6$ GeV$^2$. However, at higher $Q^2$ values, its $Q^2$ dependence becomes much softer and is consistent with the dipole parametrization assuming the axial masses 1.20--1.35 GeV. The uncertainties of the MINERvA fit are smaller than those of the $z$-expansion fit to the deuterium data~\cite{Meyer:2016oeg}.

In this analysis, we use the axial form factor from MINERvA~\cite{MINERvA:2023avz}, to check if it is consistent with the \MB{} cross sections reported in Ref.~\cite{MicroBooNE:2020fxd} within the considered nuclear models. For the SF approach, we also discuss to what extent our findings are sensitive to this choice, by performing comparisons with the results obtained using other $F_A(Q^2)$ parametrizations~\cite{Bernard:2001rs,MiniBooNE:2007iti,MiniBooNE:2010bsu,Meyer:2016oeg}.

\section{Results and discussion}\label{sec:results}
We start presentation of the \nuwro{} predictions from the cross sections for electron scattering, which are simpler to interpret than neutrino results, yet they provide important complementary information.

\begin{figure}
\centering
   \includegraphics[trim={0 0 42 0},clip,width=0.95\columnwidth]{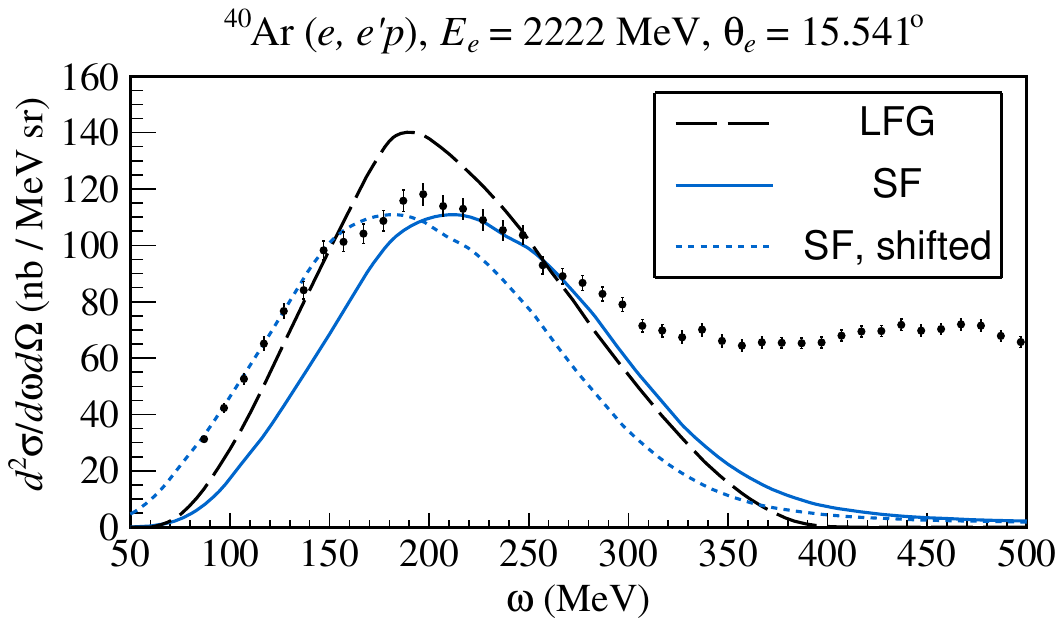}
\caption{\label{fig:electrons}Comparison of the \nuwro{} calculations of the double differential cross section for inclusive electron scattering on argon, obtained within the local Fermi gas model (dashed line) and the spectral function approach (solid line). In the latter case, the peak position differs by 30 MeV from the data, as illustrated by the shifted result (dotted line). The points represent the measurement reported by the JLab experiment E12-14-012 in Ref.~\cite{Dai:2018gch}.
}
\end{figure}

Figure~\ref{fig:electrons} compares the results obtained within the LFG model and the SF approach to the data~\cite{Dai:2018gch}, extracted for beam energy 2.222 GeV and scattering angle $15.541\degree$.\footnote{The electron-scattering cross sections are obtained from the distributions of $5\times10^9$ events with the cosine of the scattering angle differing from $\cos\theta_e$ by no more than 0.001.} The error bars of the data represent the total uncertainties. They are dominated by the (uncorrelated) statistical uncertainties, typically comprising 70\%--80\% of the total uncertainties.

The LFG calculation turns out to significantly overestimate the quasielastic peak in the region of its maximum. This behavior does not come as a~surprise, as it has been previously reported in the literature, see, e.g., Refs.~\cite{Buss:2007ar,Leitner:2008ue,Buss-thesis:2008}.

While the shape and magnitude of the SF cross section is in good agreement with the data, the position of the quasielastic peak is shifted by about 30 MeV, as shown in Fig.~\ref{fig:electrons}. This discrepancy can be ascribed to the shift introduced by FSI~\cite{Ankowski:2014yfa}, which is currently not accounted for in \nuwro{}.

We note that other mechanisms of interaction, most notably the $\Delta$ resonance excitation, start to contribute to the cross section for energy transfers above $\sim$190 MeV. Unfortunately, the accuracy of their descriptions by \nuwro{} cannot be currently tested, as they cannot be run in the electron mode.

In Ref.~\cite{MicroBooNE:2020fxd}, the \MBC{} reported various differential cross sections for an event class dubbed \ppi{}, defined according to the selection criteria listed in Appendix~\ref{sec:cuts}. To enhance the CCQE contribution, these cuts require, among others, for an event to contain a~proton and a muon, whose tracks are consistent with being coplanar with the beam axis, and missing transverse momentum is small. These kinematic selections are very efficient at removing multinucleon backgrounds, such as those originating from MEC and pion production, in a largely model-independent manner.

The \MBC{} estimated that CCQE interactions comprise $\sim$96\% of the signal \ppi{} events, using the {\sc genie} Monte Carlo generator version~\cite{MicroBooNE:2020fxd}. Using the SF approach in \nuwro{}, we find this figure to be 95.0\%. Such a good agreement suggests that the background estimates in these two simulations are consistent enough to make our comparisons to the \ppi{} data of Ref.~\cite{MicroBooNE:2020fxd} meaningful.

In a subsequent article~\cite{MicroBooNE:2023cmw}, the \MBC{} significantly relaxed the cuts, such that CCQE events contributed only $\sim$74\% of the signal events. That analysis can be expected to exhibit less reliance on the generator used for background estimation. However, comparisons to the extracted cross sections require a~complete description of all interaction channels. Therefore, here we only consider the \MB{} results of Ref.~\cite{MicroBooNE:2020fxd}, which allow us to focus on CCQE scattering.

\begin{figure}[!t]
\centering
    \includegraphics[trim={0 0 42 20},clip,width=0.95\columnwidth]{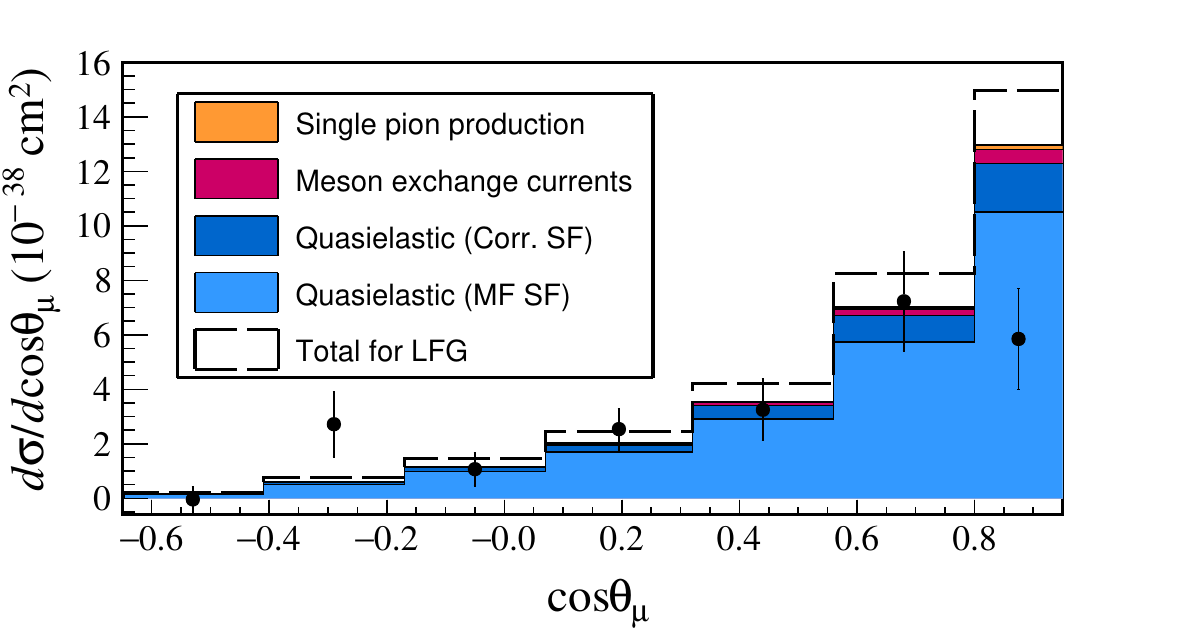}
\caption{\label{fig:nu_costheta}\MB{} \ppi{} differential cross section as a~function of the cosine of the muon production angle, calculated using the spectral function approach implemented in \nuwro{} (stacked histogram). For comparison, the local Fermi gas result (dashed-line histogram) is also shown. The experimental data are obtained from Ref.~\cite{MicroBooNE:2020fxd}.
}
\end{figure}

Figure~\ref{fig:nu_costheta} presents a~comparison of the \nuwro{} simulations, performed using the LFG model and the SF approach,\footnote{The \ppi{} cross sections for muon neutrino scattering on argon in \MB{} are obtained from the distributions of $5\times10^6$ events, $\sim$15\% ($\sim$23\%) of which pass the selection criteria detailed in Appendix~\ref{sec:cuts} for $-0.65<\cos\theta_\mu<0.80$ (0.95).} to the experimental data for the \ppi{} differential cross section as a function of the cosine of the muon production angle~\cite{MicroBooNE:2020fxd}. As shown by the dashed-line histogram, the LFG prediction generally overestimates the experimental data, with the $\chi^2/\text{d.o.f} = 5.7$ (40.0/7). This finding is consistent with what we have observed for electrons, discussing Fig.~\ref{fig:electrons}, and with the conclusions of the \MBC~\cite{MicroBooNE:2020fxd}.


In the SF approach, nuclear effects typically quench cross sections more significantly than in the LFG model. This behavior stems largely from the fact that scattering on correlated nucleons, absent in the LFG model, typically requires higher energy transfers than interactions with uncorrelated ones, which suppresses cross sections. As a consequence, the SF prediction---represented by the stacked histogram---describes the experimental data better, with $\chi^2/\text{d.o.f}=4.5$ (31.2/7).

Both the considered nuclear models conspicuously fail to reproduce the cross section for the highest $\cos\theta_\mu$ bin. However, in the subsequent analysis~\cite{MicroBooNE:2023cmw}, the \MBC{} found that this experimental point was affected by an inaccurate estimate of beam-related background and efficiency corrections. Therefore, from now on, we only discuss the results corresponding to the restricted muon phase space, $-0.65<\cos\theta_\mu<0.80$. The cross sections for the full $\cos\theta_\mu$ range are included in Appendix~\ref{sec:fullResults} for completeness.

When the highest $\cos\theta_\mu$ point is excluded from the analysis, the goodness of fit significantly improves, with the $\chi^2/\text{d.o.f}$ reducing to 1.1 (6.7/6) for the LFG model and 0.8 (5.0/6) for the SF approach.

The current \nuwro{} implementation of the SF approach does not simulate an additional nucleon in the primary interaction when the cross section is calculated using the correlated part of the SF. This issue introduces an additional source of uncertainty to our \ppi{} results, required to contain a single proton with momentum above 300 MeV.

\begin{figure}
\centering
    \includegraphics[trim={0 0 42 0},clip,width=0.95\columnwidth]{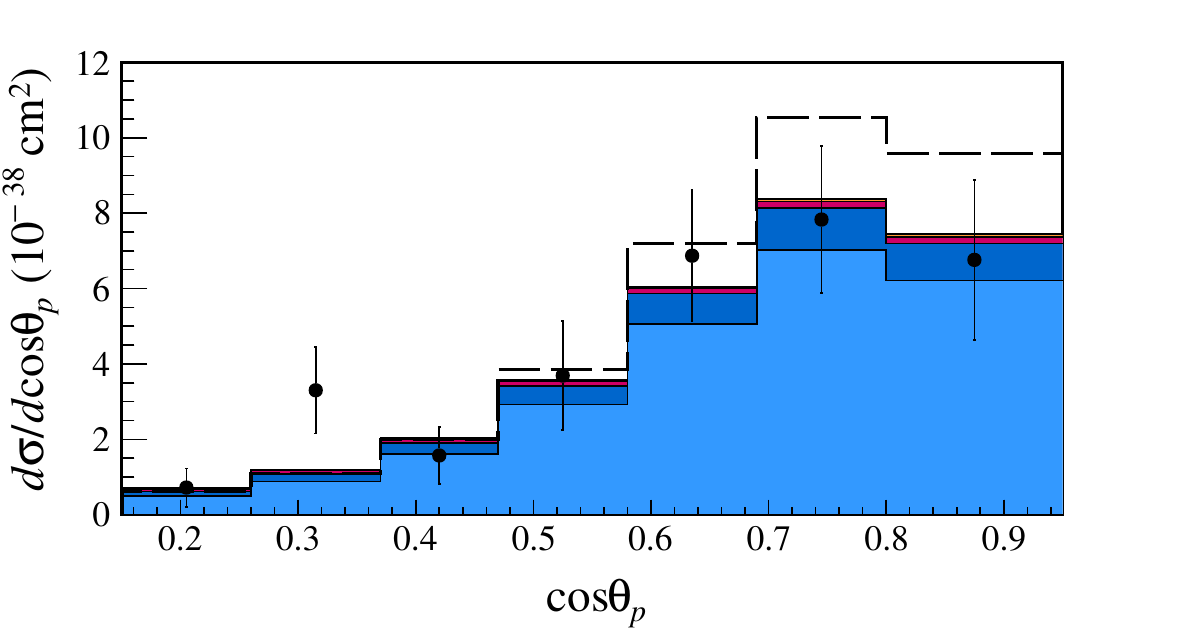}
    \includegraphics[trim={0 0 42 0},clip,width=0.95\columnwidth]{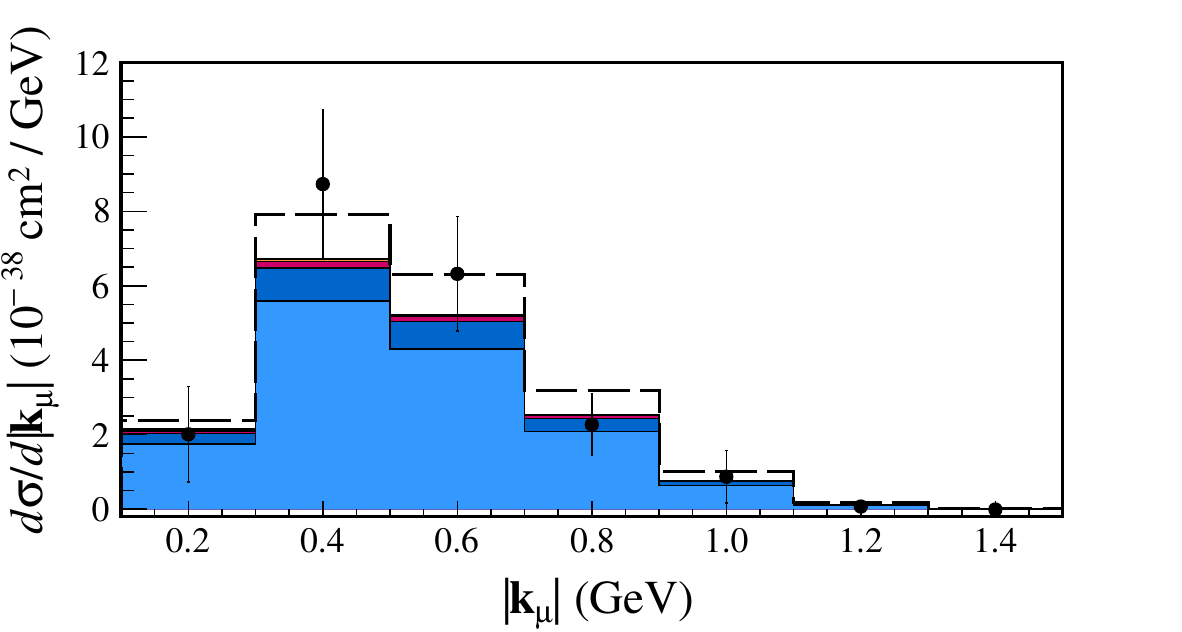}
    \includegraphics[trim={0 0 42 0},clip,width=0.95\columnwidth]{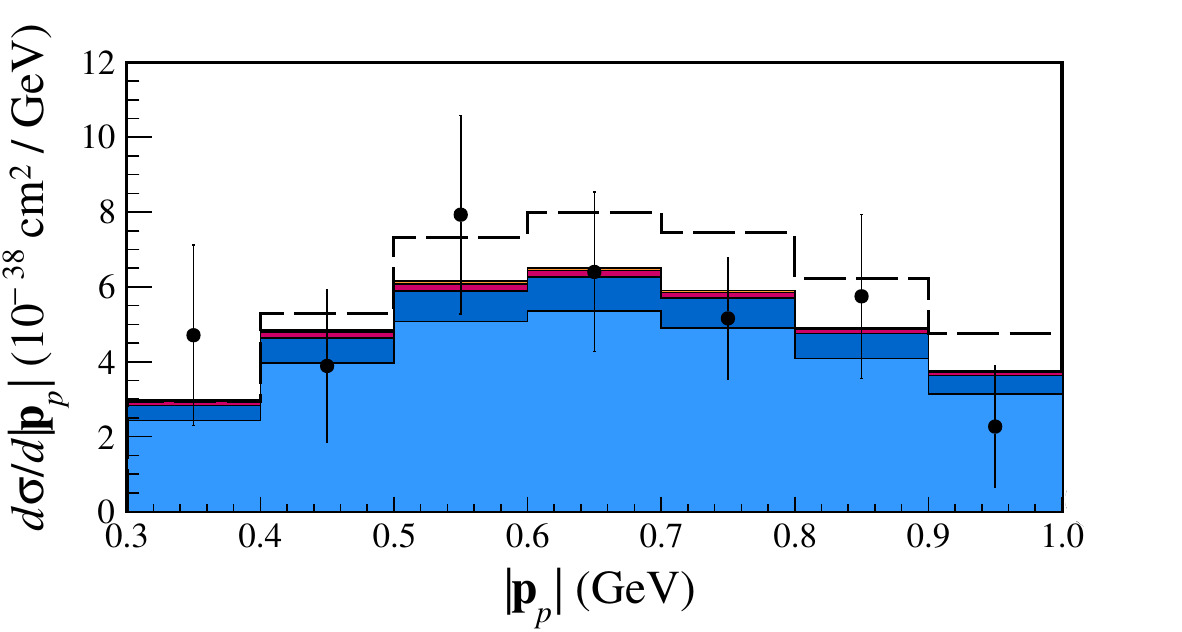}
\caption{\label{fig:nu_other}Same as Fig.~\ref{fig:nu_costheta} but for the differential cross sections as a~function of the cosine of the proton production angle (top panel), the muon momentum (middle panel), and the produced proton's momentum (bottom panel). The muon phase space is restricted, $-0.65<\cos\theta_\mu<0.80$.
}
\end{figure}

To gauge an upper limit of this uncertainty, we present the MF and correlated contributions to our \ppi{} predictions separately. It is important to note that $\sim$50\% of the correlated SF corresponds to initial momenta below 300 MeV, and FSI are expected to affect about half of the outgoing nucleons. As a consequence, we estimate the uncertainty not to exceed 25\% of the correlated contribution, which makes it not significant for the interpretation of our results, as can be seen from Figs.~\ref{fig:nu_costheta} and \ref{fig:nu_other}.

In the top panel of Fig.~\ref{fig:nu_other}, we compare the LFG and SF predictions for the cross section as a~function of the cosine of the proton production angle to the \MB{} \ppi{} data. In this case, we observe the largest difference between their $\chi^2/\text{d.o.f}$---amounting to 1.27 (8.9/7) and 0.66 (4.6/7), respectively---the dominant source of which is the shape of the theoretical results.

On the other hand, the cross sections as a function of the muon and proton momenta---shown in the middle and bottom panels of Fig.~\ref{fig:nu_other}---are reproduced by both the calculations with a similar quality of fit. The $\chi^2/\text{d.o.f}$ values range here between 0.61 and 0.77, see Table~\ref{tab:fits}, and differ predominantly due to the normalization.

\begin{table}
\caption{\label{tab:fits} The values of $\chi^2$ per degree of freedom for the agreement between the \nuwro{} simulations and the \MB{} \ppi{} data for the restricted muon phase space~\cite{MicroBooNE:2020fxd}. We compare the results for the local Fermi gas model and the spectral function approach, obtained with the MINERvA parametrization of the axial form factor~\cite{MINERvA:2023avz}.}
\begin{tabular*}{\linewidth}{@{\extracolsep{\fill}} c c c}
\toprule
 & LFG & SF \\
\midrule
$d\sigma/d\cos\theta_\mu$ & 1.11 (6.7/6)  & 0.83 (5.0/6) \\
$d\sigma/d\cos\theta_p$ & 1.27 (8.9/7)  & 0.66 (4.6/7) \\
$d\sigma/d\n{k_\mu}$ & 0.72 (5.0/7)  & 0.61 (4.3/7) \\
$d\sigma/d|{\ve p_p}|$ & 0.77 (5.4/7)  & 0.65 (4.6/7) \\
combined & 0.96 (26.0/27)  & 0.68 (18.4/27)  \\
\bottomrule
\end{tabular*}
\end{table}


\begin{table*}
\caption{\label{tab:fits_FA} Same as Table~\ref{tab:fits} but we compare different parametrizations of the axial form factor: the dipole one with the axial mass values of 1.35 GeV~\cite{MiniBooNE:2010bsu,MiniBooNE:2010xqw}, 1.23 GeV~\cite{MiniBooNE:2007iti,MINOS:2014axb}, and 1.026 GeV~\cite{Bernard:2001rs}, the $z$-expansion fit to deuterium data~\cite{Meyer:2016oeg}, and the fit to MINERvA hydrogen data~\cite{MINERvA:2023avz}. The spectral function approach is used as the nuclear model.}
\begin{tabular*}{\linewidth}{@{\extracolsep{\fill}} c c c c c c}
\toprule
 & dipole & dipole & dipole & \multirow{2}{*}{$z$-expansion} & \multirow{2}{*}{MINERvA}\\
  & $M_A=1.35$ GeV & $M_A=1.23$ GeV & $M_A=1.026$ GeV &  & \\
\midrule
$d\sigma/d\cos\theta_\mu$  & 1.18 (7.1/6) & 1.00 (6.0/6) & 0.80 (4.8/6) & 0.80 (4.8/6) & 0.83 (5.0/6) \\
$d\sigma/d\cos\theta_p$  & 1.01 (7.1/7) & 0.81 (5.7/7) & 0.65 (4.6/7) & 0.67 (4.7/7) & 0.66 (4.6/7) \\
$d\sigma/d\n{k_\mu}$ & 0.67 (4.7/7) & 0.54 (3.8/7) & 0.56 (3.9/7) & 0.62 (4.3/7) & 0.61 (4.3/7)\\
$d\sigma/d|{\ve p_p}|$ & 1.06 (7.4/7) &  0.81 (5.6/7)  & 0.62 (4.3/7) & 0.60 (4.2/7) & 0.65 (4.6/7)\\
combined & 0.97 (26.3/27) & 0.78 (21.7/27) & 0.65 (17.6/27) & 0.67 (18.0/27) & 0.68 (18.4/27)\\
\bottomrule
\end{tabular*}
\end{table*}

Note that in the approach of Ref.~\cite{Ankowski:2014yfa}, these distributions would be affected by FSI, expected to redistribute a~part of the strength toward lower (higher) values of the proton (muon) momentum. This subtle effect should lower the $\chi^2/\text{d.o.f}$ value for $d\sigma/d\n{k_\mu}$, where the uncertainties are smaller, but not for $d\sigma/d\n{p_p}$ due to its larger uncertainties.


Throughout this article, we present the cross sections obtained using the axial form factor determined by the MINERvA experiment~\cite{MINERvA:2023avz}. Let us now ponder how this choice affects the $\chi^2/\text{d.o.f}$ values, by comparing the SF results for an array of $F_A(Q^2)$ parametrizations.

\emph{A priori} the dipole form factor with the axial mass 1.20--1.35 GeV seems well motivated by the findings of MiniBooNE~\cite{MiniBooNE:2007iti,MiniBooNE:2010xqw,MiniBooNE:2010bsu}, at the same kinematics as \MB{}. On the other hand, the frequently used $M_A=1.026$ GeV~\cite{Bernard:2001rs} is partly based on the light-target data. Additionally, the $z$-expansion fit to the CCQE $\nu_\mu$ cross sections for deuterium~\cite{Meyer:2016oeg} accounts, among others, for the progress in the vector form factor determination.

As shown in Table~\ref{tab:fits_FA}, within the SF approach, the dipole parametrization with the axial mass 1.35 GeV yields the cross sections in good agreement with the \MB{} data, with $\chi^2/\text{d.o.f}\simeq1.0$. When $M_A=1.23$ GeV is used, the $\chi^2/\text{d.o.f}$ value decreases to $\sim$0.8. The three parametrizations based on the light-target data---the dipole one with the world average $M_A$ value~\cite{Bernard:2001rs}, the $z$-expansion fit to the deuterium data~\cite{Meyer:2016oeg}, and the MINERvA fit to the hydrogen cross section~\cite{MINERvA:2023avz}---describe the data with comparable quality, corresponding to $\chi^2/\text{d.o.f}\simeq0.7$.

We note that while the \MB{} kinematics is sensitive to the differences between the $F_A(Q^2)$ parametrizations, the uncertainties of the \ppi{} cross sections of Ref.~\cite{MicroBooNE:2020fxd} are too large to indicate an inconsistency with a~particular choice. The selections of the axial form factor and the nuclear model turn out to impact the goodness of fit to the same extent. Therefore, to shed light on an accurate description of CCQE scattering in argon, an updated analysis of the full dataset collected by \MB{} is urgently called for.

As a final remark, we observe that the agreement between the results of MINERvA and \MB{} was by no means guaranteed. The two experiments collected data at very different kinematic regimes---with the average energies of 0.8 GeV and 5.4 GeV---employing very different targets---hydrogen and argon---for antineutrino and neutrino interactions, and relying on different detector technologies. Therefore, the consistency between the MINERvA and \MB{} measurements can be viewed as the dawn of the precision era of neutrino experiments.

\section{Summary}\label{sec:summary}
In this article, we present a number of updates of the \nuwro{} Monte Carlo generator that improve the accuracy of the description of quasielastic interactions in the argon nucleus. We implement, among others, the spectral functions determined in a~coincidence electron-scattering experiment E12-14-012 in Jefferson Laboratory Hall A, and account for the influence of the nuclear Coulomb field on the kinematics and the wave functions of the charged particles involved in the process of scattering. These changes are of particular relevance for the search for the excess appearance events in the SBN program.

In the context of available experimental data for electron and neutrino scattering, we discuss the predictions of the updated generator, by comparing them with the results obtained using the LFG model, broadly employed by neutrino experiments.

For electrons, we illustrate both the strengths and the limitations of the updated approach, which does not currently account for the effects of FSI. For neutrinos, we analyze various differential cross sections for \ppi{} events in \MB{}~\cite{MicroBooNE:2020fxd}, and find that the spectral function approach describes the data with the $\chi^2/\text{d.o.f}$ of 0.7, compared with 1.0 for the LFG model.

The importance of our finding comes from the parametrization of the axial form factor, extracted by MINERvA from antineutrino scattering on hydrogen. Our results indicate a~consistency between \MB{} and MINERvA.

For other $F_A(Q^2)$ parametrizations partly~\cite{Bernard:2001rs} or fully~\cite{Meyer:2016oeg} based on neutrino experiments with the deuterium targets, we find comparable goodness of fits, $\chi^2/\text{d.o.f}\simeq0.7$, within the spectral function approach. Should the axial mass 1.35 GeV (1.23 GeV) be used in the dipole parametrization of the form factor, the value of $\chi^2/\text{d.o.f}$ would increase to $\sim$1.0 ($\sim$0.8), showing that the calculated cross sections remain in good agreement with the \MB{} data.

In the future developments of \nuwro{}, we plan to implement the description of FSI according to the approach of Ref.~\cite{Ankowski:2014yfa}, and separate treatments of the mean-field and correlated contributions to the spectral functions~\cite{CiofidegliAtti:1995qe}. These updates are expected to be particularly important for an accurate description of event distributions involving hadron kinematics.

It is noteworthy that according to our SF simulation, CCQE interactions comprise 96.2\% (95.0\%) of the \MB{} \ppi{} data for $\cos\theta_\mu<0.8$ ($\cos\theta_\mu<0.95$). Such a high level of purity and the relative simplicity of CCQE scattering makes them ideally suited for tests of nuclear models employed in neutrino-oscillation studies.

We strongly urge the SBN program, and the \MBC{} in particular, to provide high-statistics updates of the \ppi{} measurement in the future. This would motivate further progress in an accurate description of nuclear effects, improve the accuracy of neutrino-energy reconstruction, and maximize the sensitivity of oscillation analyses.

\begin{acknowledgments}
We would like to express our gratitude to Afroditi Papadopoulou for sharing the details of the \MB{} results, and to Tejin Cai for the enlightening correspondence on the axial form factor extraction in MINERvA. This work is partly (A.M.A., K.M.G., J.T.S.) or fully (R.D.B., B.E.K., H.P.) supported by the National Science Centre under grant UMO-2021/41/B/ST2/02778.
\end{acknowledgments}

\begin{figure}
\centering
    \includegraphics[trim={0 0 42 0},clip,width=0.95\columnwidth]{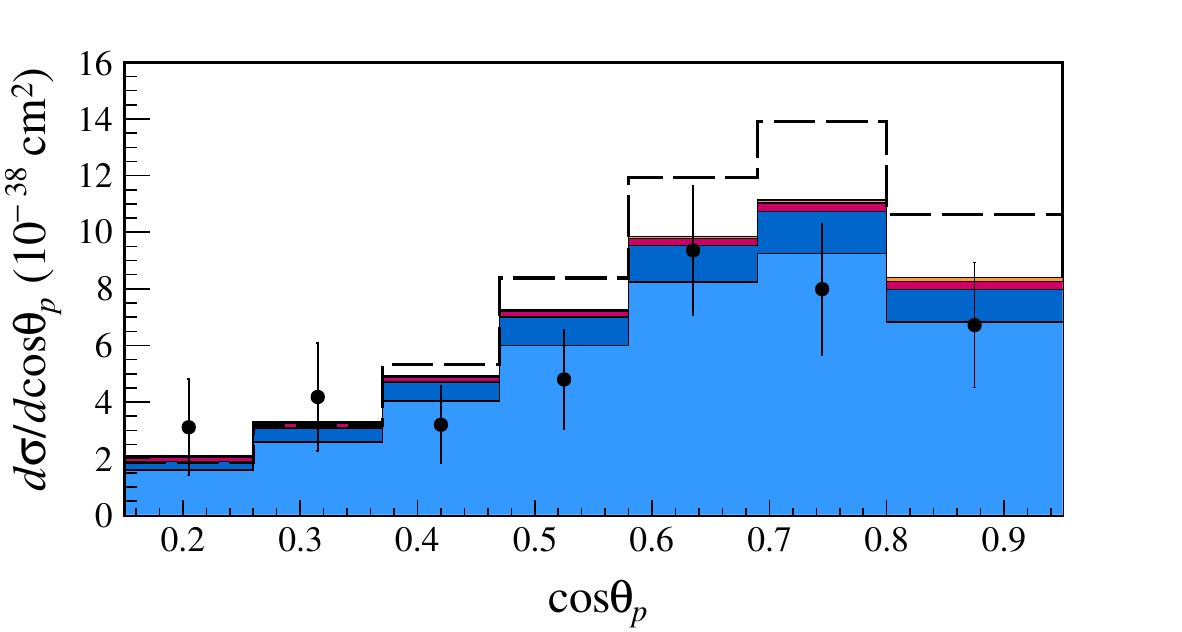}
    \includegraphics[trim={0 0 42 0},clip,width=0.95\columnwidth]{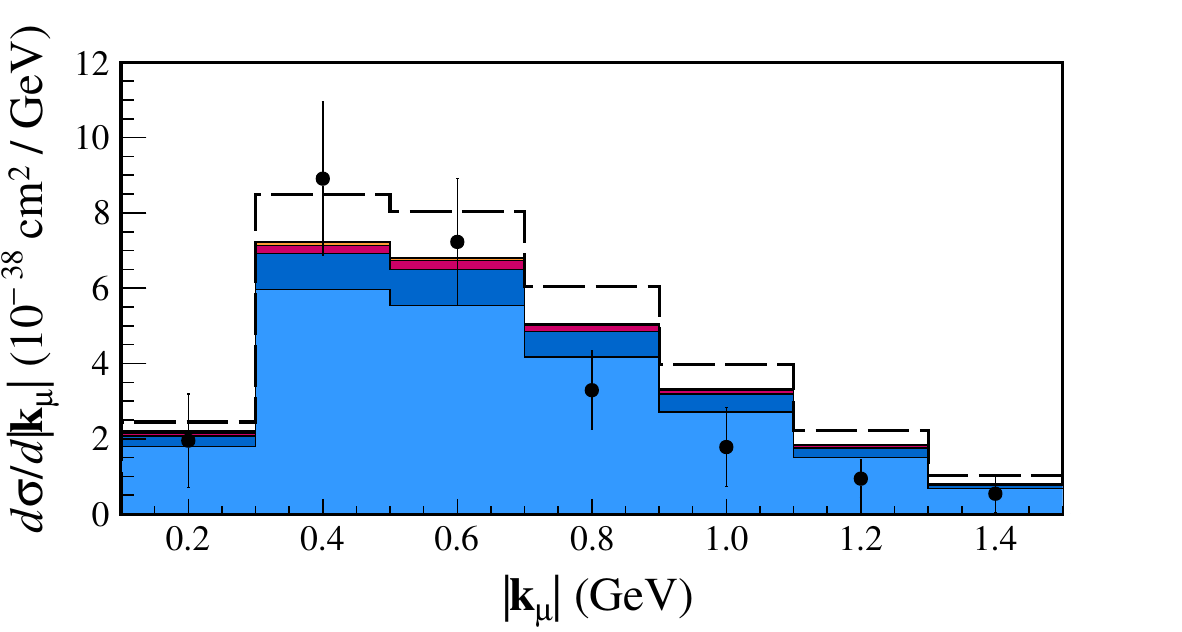}
    \includegraphics[trim={0 0 42 0},clip,width=0.95\columnwidth]{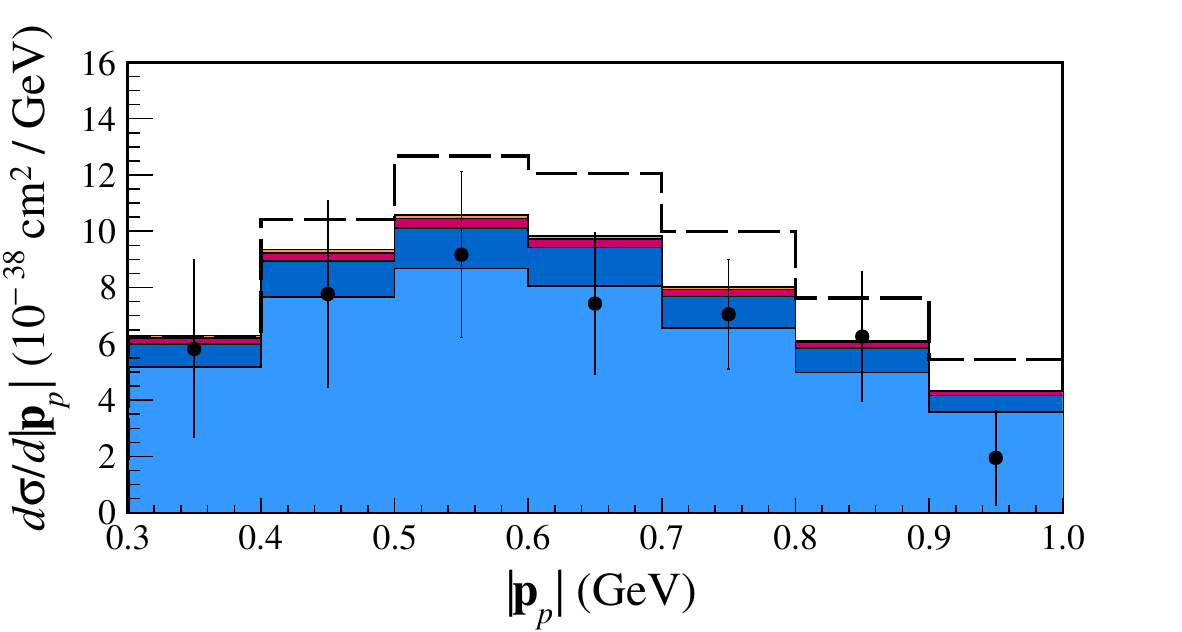}
\caption{\label{fig:nu_other_fullRange}Same as Fig.~\ref{fig:nu_costheta} but for the differential cross sections as a~function of the cosine of the proton production angle (top panel), the muon momentum (middle panel), and the produced proton's momentum (bottom panel). The muon phase space is full, $-0.65<\cos\theta_\mu<0.95$.
}
\end{figure}

\begin{table}[!b]
\caption{\label{tab:fits_fullRange} Same as Table~\ref{tab:fits} but for the full muon phase space.}
\begin{tabular*}{\linewidth}{@{\extracolsep{\fill}} c c c}
\toprule
 & LFG & SF \\
\midrule
$d\sigma/d\cos\theta_\mu$ & 5.72 (40.0/7)  & 4.46 (31.2/7) \\
$d\sigma/d\cos\theta_p$ & 1.79 (12.5/7)  & 0.87\phantom{0} (6.1/7)  \\
$d\sigma/d\n{k_\mu}$ & 2.31 (16.2/7)  & 1.53 (10.7/7)  \\
$d\sigma/d|{\ve p_p}|$ & 0.46\phantom{0} (3.2/7)  & 0.39\phantom{0} (2.7/7)  \\
combined & 2.57 (72.0/28)  & 1.81 (50.8/28)  \\
\bottomrule
\end{tabular*}
\end{table}

\appendix
\section{Employed cuts}\label{sec:cuts}
Following Refs.~\cite{MicroBooNE:2020fxd,MicroBooNE:2018vxr}, we require events to pass these selection criteria:
\begin{itemize}
\item[(a)] No $K$'s, $\eta$'s, $e^\pm$'s, $\gamma$'s, or $\pi^0$'s in the final state. No $\pi^\pm$ with momentum above 70 MeV.
\item[(b)] Leading proton of momentum $0.3< |\ve p_p|<1.0$ GeV and cosine of production angle $0.15<\cos\theta_p<0.95$. Any number of subleading protons with momenta below 0.3 GeV or above 1.0 GeV.
\item[(c)] Any number of neutrons in the final state.
\item[(d)] Single muon of momentum $0.1< \n{k_\mu}<1.5$ GeV and cosine of production angle $-0.65<\cos\theta_\mu<0.80$ (0.95) for the restricted (full) phase space.
\item[(e)] Small missing transverse momentum relative to the beam
direction, $|\ve k^T_\mu+\ve p^T_p| < 0.35$ GeV.
\item[(f)] Coplanarity of the muon's and proton's tracks, $|\Delta\phi_{\mu,p}-180\degree|<35\degree$, where
 \[\Delta\phi_{\mu,p} =\arccos\left(\frac{\ve k^T_\mu\cdot\ve p^T_p}{\n{k^T_\mu}|\ve{p}^T_p|}\right).\]
\item[(g)] Opening angle between the muon's and proton's tracks, \[\Delta\theta_{\mu,p} =\arccos\left(\frac{\ve k_\mu\cdot\ve p_p}{\n{k_\mu}|\ve{p}_p|}\right),\] restricted to
$|\Delta\theta_{\mu,p}-90\degree|<55\degree.$
\item[(h)] Muon track longer than the proton one,
\[\frac{(E_\mu-m_\mu)^3}{m_\mu}>\frac{(E_{p}-M)^3}{M}.\]
\end{itemize}

\section{Results for the full muon phase space}\label{sec:fullResults}
For the sake of completeness and to provide predictions for possible future updates of the \MB{} \ppi{} data, in Fig.~\ref{fig:nu_other_fullRange}, we present the differential cross sections for the full muon phase space, $-0.65<\cos\theta_\mu<0.95$. In Table~\ref{tab:fits_fullRange}, we also show the corresponding values of the $\chi^2$ per degree of freedom, to facilitate comparisons with other approaches published in the literature.


%

\end{document}